\pdfoutput=1 
\documentclass[prd,twocolumn,preprintnumbers,superscriptaddress,showpacs]{revtex4-1}

\usepackage{graphicx}
\usepackage{bm}
\usepackage{amsmath}
\usepackage{amssymb}
\usepackage{slashed}
\usepackage[colorlinks,pdfstartview=FitH]{hyperref}
\hypersetup{linkcolor=blue,citecolor=blue,filecolor=black,urlcolor=blue}

\def \be {\begin{equation} }
\def \ee {\end{equation}}

\def \bem {\begin{multline}}
\def \eem {\end{multline}}

\def \bes {\begin{subequations} }
\def \ees {\end{subequations}}

\def \pd {\partial}

\def \c {\chi}
\def \d {\delta}
\def \e {\epsilon}
\def \ce {\varepsilon}

\def \h {\eta}

\def \p {\pi}

\def \th {\theta}
\def \x {\xi}

\def \D {\Delta}
\def \G {\Gamma}

\def \tp {\tilde{p}}

\def \<{\langle}
\def \>{\rangle}
\def \+{\dagger}
\def \({\left(}
\def \){\right)}
\def \[{\left[}
\def \]{\right]}

\def \lo {\text{LO}}
\def \nlo {\text{NLO}}

\usepackage[normalem]{ulem}

\begin{document}

\author{Shu~Lin}
\email{linshu8@mail.sysu.edu.cn}
\affiliation{School of Physics and Astronomy, Sun Yat-Sen University, Zhuhai 519082, China}
\affiliation{Guangdong Provincial Key Laboratory of Quantum Metrology and Sensing, Sun Yat-Sen University, Zhuhai 519082, China}
\author{Yanyan Bu}
\email{yybu@hit.edu.cn}
\affiliation{School of Physics, Harbin Institute of Technology, Harbin 150001, China}
\author{Chang~Lei}
\email{leich6@mail2.sysu.edu.cn}
\affiliation{School of Physics and Astronomy, Sun Yat-Sen University, Zhuhai 519082, China}

\title{Non-Gaussianity from Schwinger-Keldysh Effective Field Theory}
\date{\today}

\begin{abstract}
We present a systematic treatment of non-Gaussianity in stochastic systems using the Schwinger-Keldysh effective field theory framework, in which the non-Gaussianity is realized as nonlinear terms in the fluctuation field. We establish two stochastic formulations of the Schwinger-Keldysh effective field theory, with those nonlinear terms manifested as multiple non-Gaussian noises in the Langevin equation and as higher order diffusive terms in the Fokker-Planck equation. The equivalence of the stochastic formulations with the original Schwinger-Keldysh effective field theory is demonstrated with non-trivial examples for arbitrary non-Gaussian parameters. The stochastic formulations will be more flexible and effective in studying non-equilibrium dynamics. We also reveal an ambiguity when coarse-graining time scale and non-Gaussian parameters vanish simultaneously, which may be responsible for the unphysical divergence found in perturbative analysis.
\end{abstract}
\maketitle

\allowdisplaybreaks

\flushbottom

\section{Introduction}
The Gaussian white noise has been widely used in modeling of stochastic dynamics. If thermodynamic limit strictly applies to the system in question, the Gaussian noise is a consequence of ensemble average due to the central limit theorem. On the other hand, the white noise follows from a coarse-grained description of the system: when the coarse-graining time scale is much longer than the microscopic time scale of the system, the white noise becomes accurate. In reality, deviations of both idealizations can occur. Non-Gaussian noises have wide applications in statistical physics \cite{BOUCHAUD1990127}, cosmology \cite{Shiraishi:2010kd}, condensed matter physics \cite{RevModPhys.86.361} and quantum optics \cite{RevModPhys.68.127}. Colored noises generically occur when one considers dynamics at time scale comparable to the coarse-graining scale. Most implementations of non-Gaussian colored noises are based on phenomenological models to date. Microscopic derivations of stochastic dynamics with non-Gaussian noises exist for a simple degree of freedom \cite{PhysRevLett.114.090601}, but generalization to more complicated systems such as hydrodynamics is far from obvious.

Modern description of a stochastic system uses Schwinger-Keldysh effective field theory (SKEFT) \cite{Kamenev2011,Crossley:2015evo,Glorioso:2017fpd,Liu:2018kfw,Haehl:2018lcu}. Thanks to doubling of degrees of freedom, the SKEFT incorporates fluctuations and dissipations systematically, going beyond the Martin-Siggia-Rose formalism for stochastic models with Gaussian noise \cite{Kamenev2011}. The SKEFT follows from averaging out fast modes and governs stochastic evolution of slow modes. The SKEFT is defined with an implicit coarse-graining time scale separating the fast modes and slow modes. The SKEFT is organized as a systematic expansion in temporal gradient, which characterizes the slowness of the dynamics, as well as expansion in the fields. These expansions allow us to study deviations of Gaussian white noise discussed above systematically: expansion in temporal gradient allows one to access dynamics comparable to the coarse-graining scale and expansion in the fields characterizes the non-Gaussianity through nonlinear effect. Recently there have been extensive studies on nonlinear effect in dynamics of Brownian particle \cite{Chakrabarty:2018dov,Chakrabarty:2019qcp,Chakrabarty:2019aeu,Jana:2021niz,Bu:2021jlp,Bu:2022oty} and hydrodynamics \cite{Chen-Lin:2018kfl,Jain:2020hcu,Jain:2020zhu}.

In this paper, we use Brownian particle as an example to illustrate formulations of non-Gaussianity from nonlinear effect. We will establish three equivalent formulations of non-Gaussianity: SKEFT, Langevin equation and Fokker-Planck (FP) equation. A crucial difference between this study and those on related subject is that we do not assume small non-Gaussian parameters and our formulations are exact in these parameters. This is in contrast to \cite{Chakrabarty:2019qcp} where equivalence has been established at lowest order in the parameters. We will offer caveat to the perturbative analysis in the non-Gaussian parameters, which contains unphysical divergence. We suggest that the divergence is tied to the ambiguity when the coarse-graining scale and non-Gaussian parameters vanish simultaneously. The formulations in this paper can also be straightforwardly adapted to more interesting hydrodynamic systems.

The rest of this paper will be structured as follows. In section \ref{SKEFT} we present a simple Schwinger-Keldysh effective field theory (EFT) incorporating non-Gaussianity. We also review its two equivalent stochastic formulations when non-Gaussian terms in noise field are absent. In section \ref{Langevin_FP} we establish two stochastic formulations of the Schwinger-Keldysh EFT when generic nonlinear terms are present. In section \ref{equivalence}, we demonstrate the equivalence of the stochastic formulations with the original Schwinger-Keldysh EFT. In section \ref{summary} we make a brief summary and outlook interesting future directions. The appendix \ref{appendix} provides further details on the equivalence demonstration.

\section{Schwinger-Keldysh effective field theory} \label{SKEFT}

We begin with the following effective Lagrangian for a Brownian particle
\begin{align}\label{Lag}
L&=iT\D_a^2-\D_a\pd_t\D_r-m\D_a\D_r+i\e_1\D_a^4-\e_2\D_a^3\D_r\nonumber\\
&+i\e_3\D_a^2\D_r^2-\e\D_a\D_r^3,
\end{align}
where $\D_r=\frac{1}{2}(\D_1+\D_2)$ and $\D_a=\D_1-\D_2$ with $\D_{1,2}$ being real scalar fields on the SK contour. $\D_r$ is identified with momentum of Brownian particle, and $\D_a$ encodes the fluctuation. We have only expanded $L$ to the leading order in temporal gradient and to quartic order in the fields. Structure like \eqref{Lag} has been obtained from holographic model calculations \cite{Bu:2021clf,Bu:2022esd}. The first three terms are Gaussian, which determine two-point correlation functions. With $T$ identified as the temperature, the first two terms satisfy the Kubo-Martin-Schwinger (KMS) symmetry \cite{Glorioso:2017fpd}
\begin{align}
&\Delta_a(t) \to -\Delta_a(-t) - iT^{-1} (\partial_t \Delta_r)(-t), \nonumber \\
&\Delta_r(t) \to - \Delta_r(-t) \label{KMS}
\end{align}
Indeed, \eqref{KMS} leaves the third and last terms invariant up to a total derivative. The remaining non-Gaussian terms determine higher-point correlation functions. One may further constrain these terms using KMS symmetry 
, which amounts to choosing an equilibrium state \cite{Jain:2020hcu}. Nevertheless, we choose to not impose the KMS symmetry for the non-Gaussian terms, which is applicable to a non-equilibrium state. Indeed, model calculations have shown violation of the KMS symmetry in higher-point correlation functions from non-Gaussian terms once the condition of equilibrium state is relaxed \cite{Bu:2021jlp}. 
All nonlinear parameters $\epsilon_{1,2,3}, \epsilon$ are real by $Z_2$-reflection symmetry of SKEFT \cite{Glorioso:2017fpd}.

Eq.~\eqref{Lag} can be inspected as a series expansion in $\D_a$. At linear order, the action variation with respect to $\D_a$ gives the deterministic equation for $\D_r$:
\begin{align}\label{det_eq}
\pd_t\D_r=-m\D_r-\e\D_r^3.
\end{align}
We readily identify \eqref{det_eq} as a non-linear damping equation of particle's momentum. Stability requires $m>0$ and $\e>0$ so that the corresponding terms act like restoring force when $\D_r$ moves away from the origin.
The quadratic terms in $\D_a$ encode stochastic property of the system, turning the deterministic equation \eqref{det_eq} into a stochastic one with Gaussian noise.
As we shall show, the remaining cubic and quartic terms in $\D_a$ give rise to non-Gaussian noises. For completeness, we will review the derivation of two well-known formulations of stochastic dynamics: Langevin equation and FP equation from the Gaussian terms (linear and quadratic in $\D_a$). Then we will extend the analysis by including non-Gaussian terms (cubic and quartic in $\D_a$).

Before proceeding, we remark that \eqref{Lag} also contains nonlinearity in $\D_r$. However, this nonlinearity exists generically in interacting systems without stochasticity, thus not affecting the noise. It is well-known how to treat this with standard perturbative method. So the non-Gaussianity inherent to stochastic systems arises from the nonlinear $\epsilon_{1,2}$-terms.


We start by converting the first term in \eqref{Lag} into a noise term added to \eqref{det_eq}. Following standard procedure, we rewrite the first term in the path integral as \cite{Kamenev2011}
\begin{align}\label{Gaussian}
e^{-\int dt T\D_a^2}=\int {\cal D}\x e^{-\int dt\(\frac{\x^2}{4T}-i\x \D_a\)}.
\end{align}
With $\x$ introduced, it is easy to integrate out $\D_a$
\begin{align}
&\int {\cal D}\D_a e^{i \int dt L}=\int {\cal D}\x e^{-\int dt\frac{\x^2}{4T}}\d(-\pd_t\D_r-m\D_r\nonumber\\
&-\e\D_r^3+\x), \nonumber
\end{align}
which gives rise to the following Langevin equation with a non-linear damping term
\begin{align}\label{Langevin}
\pd_t\D_r=-m\D_r-\e\D_r^3+\x.
\end{align}
$\x$ is identified as a Gaussian noise whose variance is determined by the exponent $e^{-\int dt\frac{\x^2}{4T}}$ as
\begin{align}\label{noise}
\<\x\>=0,\quad\<\x(t)\x(t')\>=2T\d(t-t').
\end{align}

The $\e_3$-term can be included straightforwardly, which turns \eqref{Langevin} into a multiplicative form
\begin{align}\label{Langevin_mul}
\pd_t\D_r=-m\D_r-\e\D_r^3+(1+\e_3 T^{-1}\D_{r}^2)^{1/2}\x.
\end{align}
We require $\e_3>0$ so that prefactor of the noise is real \footnote{If $\e_3<0$, the counterpart of \eqref{Gaussian} diverges leading to an unstable theory}.
It is well-known that \eqref{Langevin_mul} with \eqref{noise} is ambiguous. We will adopt the Ito-regularization \cite{Kamenev2011}
\begin{align}\label{ito}
&\D_{r,i}-\D_{r,i-1}=\d_t[-m\D_{r,i-1}-\e\D_{r,i-1}^3\nonumber\\
&+(1+\e_3 T^{-1}\D_{r,i-1}^2)^{1/2}\x_i],
\end{align}
in which the multiplicative factor of noise at step $i$ depends on the field at one step earlier. The noise is normalized as $\<\x_i\x_j\>=2T\d_{ij}\d_t^{-1}$ \footnote{$\d_t$ is absorbed into the definition of the discretized noise by convention.}, with $\x_i$ being the discretized noise at step $i$. $\d_t$ is the timestep used in discretization, which is also the coarse-graining time scale.

It is convenient to pass from SKEFT to the FP equation, which governs the evolution of probability function $P(t,\D_r)$. The probability satisfies the following evolution equation from the discretized path integral
\begin{align}\label{FP_L}
&P(t_{i},\D_{r,i})=\int d\D_{r,i-1}d\D_{a,i}{\rm exp} \big[-\d_tT\D_{a,i}^2-i\D_{a,i}\d_r\nonumber\\
&-i\d_tm\D_{a,i}\D_{r,i-1}-\d_t\e_3\D_{a,i}^2\D_{r,i-1}^2-i\d_t\e\D_{a,i}\D_{r,i-1}^3\big]\nonumber\\
&P(t_{i-1},\D_{r,i-1}),
\end{align}
with $\d_r=\D_{r,i}-\D_{r,i-1}$. In the limit $\d_t\to0$, the exponent suggests $\D_{a,i}\sim\d_t^{-1/2}$ and $\d_r\sim\d_t^{1/2}$. We may regard $t$ and $\D_r$ as continuous variables and expand
\begin{align}\label{expansion}
&P(t_{i-1},\D_{r,{i-1}})=P(t_{i},\D_{r,{i}})-\d_t\dot{P}(t_{i},\D_{r,i})\nonumber\\
&-\d_rP'(t_{i},\D_{r,i})+\frac{1}{2}\d_r^2P''(t_{i},\D_{r,i})+\cdots,
\end{align}
with dot and prime denoting derivatives with respect to $t$ and $\D_r$ respectively. Plugging \eqref{expansion} into \eqref{FP_L} and making a change of field $d\D_{r,i-1}=d\d_r$, we can perform the integrals easily to obtain the following FP equation from the coefficient of $\d_t$.
\begin{align}
\pd_tP= & T\pd_{\D}^2P + m\pd_{\D}(\D P)+\e_3\pd_{\D}^2(\D^2P) + \e\pd_{\D}(\D^3P).
\end{align}
We have renamed $\D_r\to\D$ for notational simplicity. Here, in accord with \eqref{ito}, Ito-regularization has been assumed so that $\pd_{\D}$ is always later than $\D$.

\section{Non-Gaussian Langevin and FP equations} \label{Langevin_FP}
So far what has been presented is textbook materials \cite{Kamenev2011}. Now we wish to generalize the Langevin and FP equations to the non-Gaussian case. Naive application of the method outlined above encounters immediate difficulties. The derivation of the Langevin equation relies on Gaussian integration, which cannot treat cubic and quartic terms in $\D_a$; the derivation of FP equation seems to involve potential divergence. If we still choose $\D_a\sim\d_t^{-1/2}$ as in the Gaussian case, the term $i\e_1\D_a^4$ for example becomes singular in the continuum limit $\d_t\to0$.

The difficulties associated with the two equations are in fact related: the technical difficulty of the non-Gaussian integration is tied to the fact that there is no simple scaling of $\D_a$ with $\d_t$ in the multi-scale integral, thus we would not have a simple noise with $\x\sim\d_t^{-1/2}$ like in the Gaussian case. Similarly, if we were able to perform the non-Gaussian integral based on \eqref{Lag} in full, i.e., without assuming a simple scaling of $\D_a$, we would not expect any divergence. Indeed the Lagrangian \eqref{Lag} is essentially a quantum mechanical one with all the couplings having mass dimension one, so the SKEFT is super-renormalizable.

The analysis above suggests that we should treat nonlinear term separately rather than assume a uniform scaling. Below we shall derive the non-Gaussian Langevin and FP equations and demonstrate their equivalence with the SKEFT formulation.

Let us begin with the non-Gaussian Langevin equation. Note that \eqref{Gaussian} for an infinitesimal time interval reads,
\begin{align}
e^{-\d_tT\D_{a,i}^2}=\int d\x_i e^{i \d_t \x_i \D_{a,i}} e^{-\d_t \frac{\x_i^2}{4T}},
\end{align}
which allows us to trade $\D_{a,i}$ with $\x_i$. Note that this is nothing but an inverse Fourier transform. We can apply the same transform to the cubic and quartic terms (omitting subscript $i$ for notational simplicity)
\begin{align}\label{nonGaussian}
&e^{-\d_t\e_1\D_a^4}=\int d\h e^{i\d_t \h \D_a}f(\h),\nonumber\\
&e^{-i\d_t\e_2\D_a^3\D_r}=\int d\c e^{i\d_t \c \D_a}g(\c),
\end{align}
with
\begin{align}\label{weights}
f(\h)= &\frac{1}{2\p}\bigg[ 2\(\frac{\d_t^3}{\e_1}\)^{1/4}\G\(\frac{5}{4}\){}_0F_2\(\frac{1}{2},\frac{3}{4};\frac{\h^4\d_t^3}{256\e_1}\) -\(\frac{\h^2}{4}\)\nonumber \\
&\times\(\frac{\d_t^3}{\e_1}\)^{3/4}  \G\(\frac{3}{4}\)  {}_0F_2 \(\frac{3}{2},\frac{5}{4};\frac{\h^4\d_t^3}{256\e_1}\)\bigg], \\
g(\c)=&\(\frac{\d_t^2}{3\e_2\D_r}\)^{1/3}Ai\(-\c\(\frac{\d_t^2}{3\e_2\D_r}\)^{1/3}\). \nonumber
\end{align}
Here ${}_0F_2$ and $Ai$ are generalized hypergeometric function and Airy function respectively. As in the Gaussian case, $f(\h)$ and $g(\c)$ are interpreted as weight of noises $\h$ and $\c$ respectively.

Note that the non-Gaussian parameters appear in the weight functions as $\d_t^3/\e_1$ and $\d_t^2/\e_2$, which are ambiguous in the limits $\d_t\to0$ and $\e_{1,2}\to0$. We suggest that the unphysical divergence found in earlier studies may be due to improper implementation of the limit in perturbative analysis in the continuum form \cite{Chakrabarty:2019qcp,Jana:2021niz}. Our derivation indicates that it is crucial to proceed in discretized form and not to use perturbation.
We also note that $\h$ and $\c$ scale differently with $\d_t$: $\h\sim \d_t^{-3/4}$, $\c\sim \d_t^{-2/3}$.

\begin{figure}[htbp]
     \begin{center}
          \includegraphics[width=0.49\textwidth]{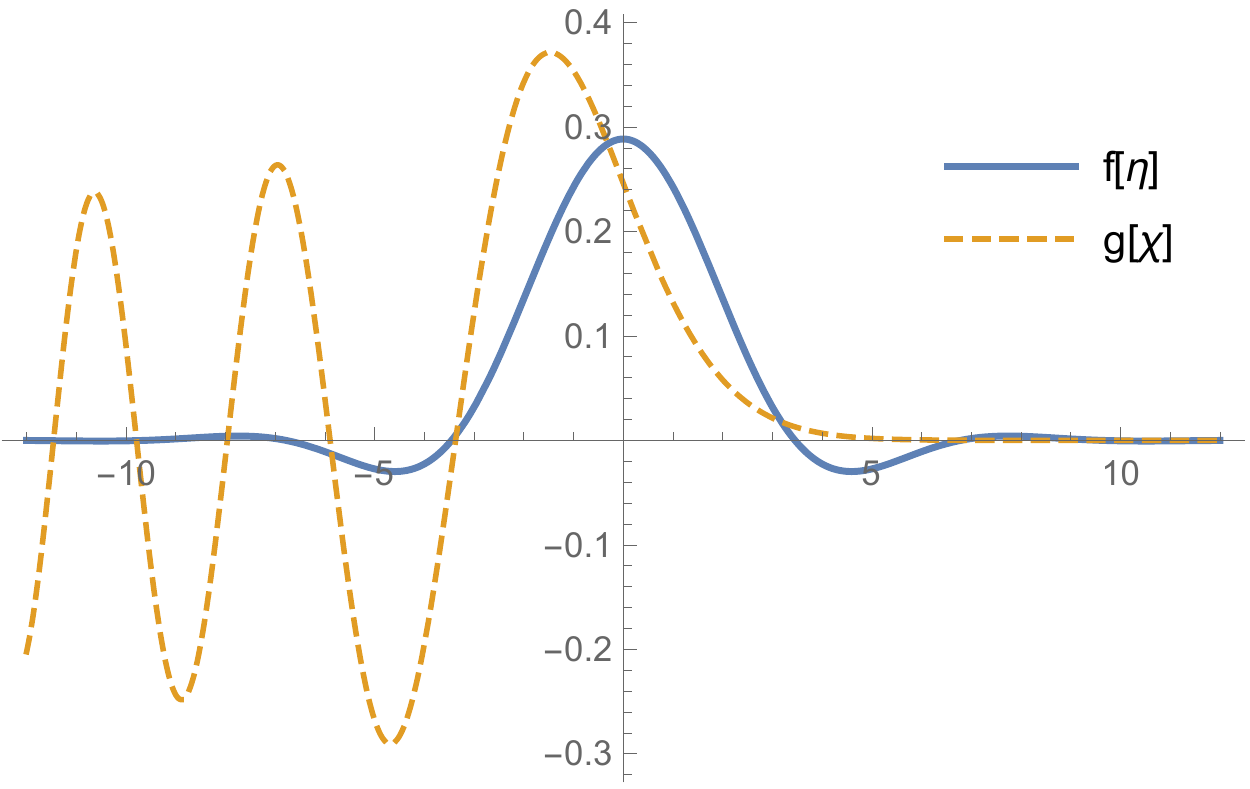}
          \caption{Non-positive definite weight functions $f(\h)$ and $g(\c)$, with $\epsilon_1/\delta_t^3 = \epsilon_2 \D_r/\delta_t^2 =1$.}
    \label{fig:weight}
    \end{center}
\end{figure}

The weight functions are plotted in Fig.~\ref{fig:weight}. Two comments of the weight functions are in order. Firstly, they are not positive definite. This is unavoidable: by taking derivatives with respect to $\D_a$ in \eqref{nonGaussian} and then setting $\D_a=0$, we can show the first nonvanishing moments are $\<\h^4\>$ and $\<\c^3\>$ respectively \footnote{In doing so, we need to make sure the integrand tends to zero at infinities. This is not satisfied by $g(\c)$. We need to introduce a regulator $e^{\pm \ce\c}$ with the plus/minus sign for $\D_r>0$ and $\D_r<0$ respectively.}. Therefore, region with negative weight must be present such that $\<\h^2\>=\<\c^2\>=0$. Secondly, $g(\c)$ depends on $\D_r$. In the Ito regularization scheme, the weight of $\c_i$ depends on $\D_{r,i-1}$. In the special limit $\D_r\to0$, we can see from the definition \eqref{nonGaussian} that $g(\c)\to\d(\c)$.

It is then straightforward to integrate out $\D_{a,i}$ to arrive at the Langevin equation with multiple noises
\begin{align}\label{L_nonG}
&\D_{r,i}-\D_{r,i-1}=\d_t \left[ -m\D_{r,i-1}-\e\D_{r,i-1}^3 \right. \nonumber\\
&\left. +(1+\e_3 T^{-1} \D_{r,i-1}^2)^{1/2}\x_i + \h_i+\c_i \right].
\end{align}

Now we turn to the derivation of the FP equation. Note that in \eqref{Lag} only $\D_r$ is the dynamical field and $\D_a$ is an auxiliary one that we wish to integrate out. By performing the Legendre transformation, we find $\D_a$ and $\D_r$ form a conjugate pair
\begin{align}
p=\frac{\pd L}{\pd( \pd_t\D_r)}=-\D_a.
\end{align}
The corresponding Hamiltonian is expressed as
\begin{align}
H=&-iT\D_a^2+m\D_a\D_r-i\e_1\D_a^4+\e_2\D_a^3\D_r\nonumber\\
&-i\e_3\D_a^2\D_r^2+\e\D_a\D_r^3.
\end{align}
In order to describe evolution of probability distribution, we need to promote the classical Hamiltonian to a quantum one, in which the conjugate fields have the commutator $[p,\D_r]=-i$. This allows us to represent the operator $p=-i\pd_{\D_r}$. It is convenient to use $\tilde{p}=-ip$. Then, the Hamiltonian appears purely imaginary
\begin{align}
&H=i\(T\tp^2-m\tp\D_r-\e_1\tp^4+\e_2\tp^3\D_r+\e_3\tp^2\D_r^2-\e\tp\D_r^3\).
\end{align}

Recall in the Gaussian case, the probability distribution $P$ evolves according to the phase factor $e^{i\int dtL}$, which is equivalent to the Schr\"odinger equation $i\pd_tP=HP$. However, in the non-Gaussian case, the equivalence is lost due to cubic and quartic terms in $\tp$. We should resort to the Schr\"odinger equation for the evolution. Upon using $p=-i\pd_\D$, we have
\begin{align}\label{FP_H}
&\pd_tP=T\pd_\D^2P+m\pd_\D(\D P)-\e_1\pd_\D^4P-\e_2\pd_\D^3(\D P)\nonumber\\
&+\e_3\pd_\D^2(\D^2P)+\e\pd_\D(\D^3P).
\end{align}
We have renamed $\D_r\to\D$ for notational simplicity. Again the ordering of operator $\pd_\D$ and $\D$ matters. We have adopted the Ito regularization so that $\pd_\D$ is always later than $\D$, to be consistent with the regularization in Langevin equation. \eqref{FP_H} generalizes \eqref{FP_L} to the non-Gaussian case. Here the term $-\e_1\pd_\D^4P$ can be viewed as a higher order diffusive term in addition to the diffusive term $T\pd_\D^2P$. Stability of the diffusion requires $\e_1>0$. The sign of $\e_2$ is not constrained.

\section{Equivalence Demonstration} \label{equivalence}
Now we demonstrate the equivalence of the three formulations by calculating a same set of equal-time correlation functions. As simple examples, we consider $\<\D^2(t)\>$ and $\<\D^4(t)\>_c\equiv\<\D^4(t)\>-3\<\D^2(t)\>^2$, in which the latter is the connected part of the 4-point correlation function. These correlation functions are directly comparable among the three formulations. We have argued below \eqref{det_eq}, \eqref{Langevin_mul} and \eqref{FP_H} that all parameters except $\e_2$ are constrained to be positive.

Let us first calculate them by solving the FP equation \eqref{FP_H}. Multiplying $d\D\D^2$ on both sides of \eqref{FP_H} and integrating by parts, we obtain the following equation for the 2nd moment
\begin{align}\label{2nd}
\pd_t\<\D^2\>=2T-2m\<\D^2\>+2\e_3\<\D^2\>-2\e\<\D^4\>.
\end{align}
The equation for $\<\D^2\>$ does not close as it involves $\<\D^4\>$. This is of course allowed, reflecting effect of nonlinearity in $\D_r$. Since our focus is on nonlinear terms in $\D_a$, we will set $\e=0$ to simplify the comparison. Then, the equation for the 4th moment reads
\begin{align}\label{4th}
\pd_t\<\D^4\>=&12(T+2\epsilon_2)\<\D^2\>-24\e_1+4(3\e_3-m)\<\D^4\>.
\end{align}

From \eqref{2nd} and \eqref{4th}, we see that in order for the moments not to blow up, we need $m>\e_3$ and $m>3\e_3$. In fact, we can obtain the condition for the $2n$-th moment would be $m>(2n-1)\e_3$, which will eventually fail for sufficient large $n$. The reason for the failure can be seen from the Langevin equation \eqref{L_nonG}: for large $\D$, the multiplicative noise always win over the $m$-term, but it can be cured by the $\e$-term we choose to turn off. Consequently to ensure stability of the FP equation, we should set $\e_3=0$ in \eqref{2nd} and \eqref{4th} as well.
This leaves us with $\e_1$ and $\e_2$ being the only nonvanishing parameters. We stress that they are also the full non-Gaussian parameters in the SKEFT.

In fact, we can show $\e_2$ is also constrained to be positive from the equation for the $2n$-th moment
\begin{align}
&\pd_t\<\D^{2n}\>=2n(2n-1)\(T+\e_2(2n-2)\)\<\D^{2n-2}\>\nonumber\\
&-2nm\<\D^{2n}\>-2n(2n-1)(2n-2)(2n-3)\e_1\<\D^{2n-4}\>.
\end{align}
Since all even moments are positive, a negative $\e_2$ combined with the other two terms on the right hand side implies a negative $\pd_t\<\D^{2n}\>$ for large enough $n$. Consequently the final state can only have vanishing moments for large $n$. We do not consider this trivial possibility. Below we take all parameters to be positive.

To solve \eqref{2nd} and \eqref{4th}, we need to specify initial condition \footnote{Vanishing boundary conditions at $\D\to\pm\infty$ have been implemented in the procedure above already}. In fact, as $t\to + \infty$, stability of the FP equation ensures that any initial state will approach the unique steady state, with all the moments tending to constants. We stress that the steady state reached in the presence of non-Gaussian noise is non-equilibrium in the absence of KMS condition, see also \cite{PhysRevE.87.052124,levy} for examples in phenomenological models. With this in mind, we take initial condition $\langle \D^2(0)\rangle = \langle \D^4(0) \rangle =0$ and easily obtain
\begin{align}\label{FP_moments}
&\<\D^2(t)\>=\frac{T}{m}\left( 1- e^{-2mt} \right),\nonumber \\
&\<\D^4(t)\>_c=-\frac{6\e_1}{m} (1-e^{-4mt})+\frac{6T\e_2}{m^2}(1-e^{-2mt})^2.
\end{align}
It is worth pointing out that \eqref{FP_moments} is exact in the non-Gaussian parameters $\e_1$ and $\e_2$.

%
%
Now we attempt to solve the Langevin equation \eqref{L_nonG}. The usual method is to generate an ensemble of solutions to Langevin equation with random noises and then take the ensemble average. There is no conceptual difficulty with this method. The issue of negative weight can be treated with technique like in \cite{PhysRevB.37.5024}. However, the introduction of the multiple noises makes the practical computation expensive. For the purpose of demonstrating the equivalence, we will use a hybrid method, in which the ensemble average is taken analytically and we only simulate the moments equation derived from the Langevin equation \eqref{L_nonG}.

From \eqref{L_nonG}, we easily obtain
\begin{align}\label{L_moments}
&\<\D_{i+1}^2\>=(1-m\d_t)^2\<\D_i^2\>+\<\x_i^2\>\d_t^2,\nonumber\\
&\<\D_{i+1}^4\>=(1-m\d_t)^4\<\D_i^4\>+6(1-m\d_t)^2\d_t^2\<\x_i^2\>\<\D_i^2\>\nonumber\\
&+\d_t^4\<\x_i^4\>+4(1-m\d_t)\d_t^3\<\D_i\c_i^3\>+\<\h_i^4\>\d_t^4,
\end{align}
where various noise averages are \footnote{The negative $4$th-moment of $\h_i$ is allowed by the non-positive definite weight function $f(\h)$.}
\begin{align}\label{noise_avg}
&\<\x_i^2\>=2T/\d_t,\quad \<\x_i^4\>=12T^2/\d_t^2,\quad \<\h_i^4\>=-24\e_1/\d_t^3\nonumber\\
&\<\c_i^3\>=6\e_2\D_i/\d_t^2\;\Rightarrow\; \<\D_i\c_i^3\>=6\e_2\<\D_i^2\>/\d_t^2.
\end{align}
Then, it is easy to show that \eqref{L_moments} leads to
\begin{align}\label{moments_dt}
&\frac{\<\D_{i+1}^2\>-\<\D_{i}^2\>}{\d_t}=-2m\<\D_i^2\>+2T+O(\d_t),\nonumber\\
&\frac{\<\D_{i+1}^4\>-\<\D_{i}^4\>}{\d_t}=-4m\<\D_i^2\>+12T\<\D_i^2\>+2\e_2\<\D_i^2\>-24\e_1\nonumber\\
&+O(\d_t).
\end{align}
which are obviously discretized forms of \eqref{2nd} and \eqref{4th} (with $\e_3=\e=0$ as we have reasoned). Therefore, we are guaranteed to arrive at the same results as \eqref{FP_moments} in the limit $\d_t\to0$.

Finally, we calculate the same quantities within the SKEFT. Note that the effective Lagrangian in the absence of external sources corresponds to the steady state discussed above, which is reached by evolving the Langevin and FP equations till the limit $t\to\infty$. Then, $\<\D(t)^2\>$ and $\<\D(t)^4\>_c$ are represented by $\<\D_r(t)^2\>$ and $\<\D_r(t)^4\>_c$, which are most conveniently calculated diagrammatically in the $ra$ basis. We treat the first three terms in \eqref{Lag} as the free part, giving the following propagators
\begin{align}\label{propagators}
&D_{rr}(t,t')=\frac{T}{m}e^{-m|t-t'|},\;D_{ra}(t,t')=-ie^{-m(t-t')}\th(t-t'),\nonumber\\
&D_{ar}(t,t')=D_{ra}(t',t).
\end{align}
The remaining terms give two vertices $-4!\e_1$ and $-4!i\e_2$.

\begin{figure}[htbp]
     \begin{center}
          \includegraphics[width=0.35\textwidth]{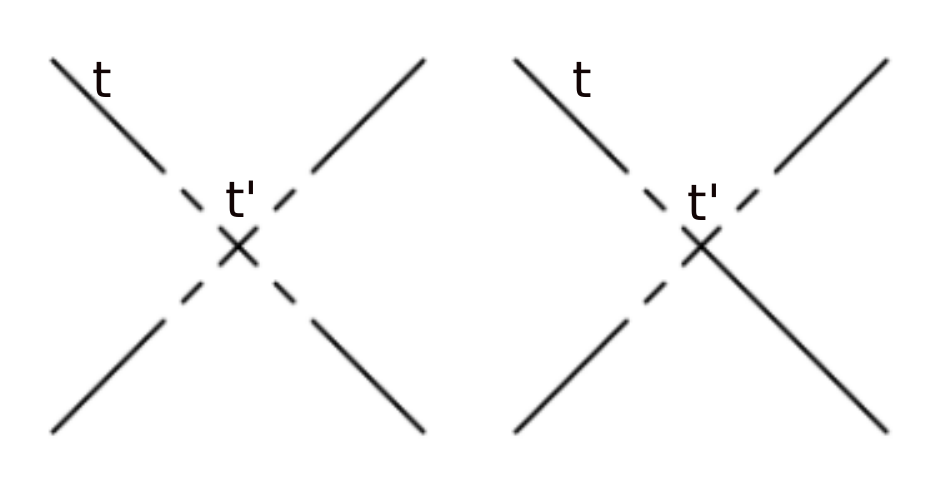}
          \caption{Feynman diagrams for $\langle \D_r(t)^4\rangle_c$ when $\epsilon_3 = \epsilon =0$. The ends of propagator can be of either dashed type or solid type, corresponding to $a$ and $r$ indices respectively.}
    \label{fig:diagrams}
    \end{center}
\end{figure}

The two-point correlation function is trivially given by the symmetric propagator $\<\D_r(t)\D_r(t)\>=\frac{T}{m}$.
The connected four-point correlation function receives contribution from diagrams in Fig.~\ref{fig:diagrams}.
Note that possible loop diagrams vanish identically because a loop involves only retarded (or advanced) propagators. The diagrams in Fig.~\ref{fig:diagrams} are easily evaluated to give
\begin{align}\label{tree}
\<\D_r(t)^4\>_c=&\int_{-\infty}^tdt'D_{ra}(t,t')^4(-4!\e_1)\nonumber\\
&+\int_{-\infty}^tdt'D_{ra}(t,t')^3D_{rr}(t,t')(-4!i\e_2)\nonumber\\
=&-\frac{6\e_1}{m}+\frac{6T\e_2}{m^2}.
\end{align}
\eqref{tree} is in full agreement with \eqref{FP_moments} in the limit $t\to\infty$.

To further corroborate the equivalence, we have also performed a more sophisticated demonstration in Appendix, in which we turn on both $\e$ and $\e_3$. Performing calculations perturbatively in these two parameters, we still find perfect agreement among three formulations with
\begin{align}\label{supp}
&\<\D_r(t)^2\>=\frac{T}{m}-\frac{3T^2\e}{m^3}+\frac{T\e_3}{m^2}+\frac{6\e_1\e}{m^2}-\frac{6T\e_2\e}{m^3},\nonumber\\
&\<\D_r(t)^4\>_c=-\frac{6\e_1}{m}+\frac{6T\e_2}{m^2}
+\frac{6(T^2-3m\e_1+4T\e_2)\e_3}{m^3}\nonumber\\
&
-\frac{6\(T^3-12Tm\e_1+15T^2\e_2-26m\e_1\e_2+26T\e_2^2\)\e}{m^4}.
\end{align}
\eqref{supp} contains extra contributions at $O(\e)$ and $O(\e_3)$. In field theoretic approach, those contributions arise from about thirty loop diagrams, while in stochastic approaches, they are obtained simply by solving the FP and Langevin equations perturbatively in $\e$ and $\e_3$. Thus the stochastic formulations provide a very efficient method of resumming diagrams, a notable virtue of the stochastic approaches.

\section{Summary and Outlook} \label{summary}
We studied effect of nonlinear terms in a SKEFT expanded up to quartic order in the fields. There are nonlinearities in both the vev field $\D_r$ and the fluctuation field $\D_a$. The nonlinearity in $\D_r$ is not specific to stochastic system and is known how to treat perturbatively. Our emphasis is on the cubic and quartic terms of $\D_a$, which are generically present in a complete theory but usually ignored in literature. We established two stochastic formulations of the SKEFT: non-Gaussian Langevin equation and FP equation. In the former case, the cubic and quartic terms of $\D_a$ are manifested as two non-Gaussian noises in addition to the Gaussian noise corresponding to the quadratic terms of $\D_a$. These two non-Gaussian noises have distinct scalings with the timestep $\d_t$ of the discrete Langevin equation. In the latter case, the nonlinear terms are manifested as higher order diffusive terms in the FP equation. Our formulations reveal an ambiguity as $\d_t$ and non-Gaussian parameters tend to zero simultaneously, shedding light on the origin of unphysical divergence found in early studies. We demonstrated the equivalence among the three formulations for arbitrary non-Gaussian parameters subject to stability conditions. 

The stochastic formulations established in this work will be found useful in addressing non-equilibrium dynamics with more flexibility and efficiency. On the one hand, it allows one to study non-equilibrium states by simply changing initial conditions. On the other hand, numerical implementation will help to efficiently resum diagrams in field theoretic approach. Moreover, by solving the non-Gaussian Langevin or FP equations, we could obtain unequal-time correlators \cite{Chakrabarty:2018dov,Chakrabarty:2019qcp,Moreau:2019jpn}, which would contain more important information about non-equilibrium physics.

The method outlined in this work can be generalized to more interesting theories such as hydrodynamics. Rapid progresses have been made in studying the effect of nonlinear Gaussian terms (quadratic in fluctuation field) \cite{Chen-Lin:2018kfl,Jain:2020hcu}. Effect of non-Gaussian terms has also been discussed recently \cite{Jain:2020zhu}. It would be useful to implement stochastic hydrodynamics with non-Gaussian noise, which would allow us to study a complete hydrodynamics out-of-equilibrium.

\section*{Acknowledgments}

This work is in part supported by NSFC under Grant Nos 12075328 and 11735007 (S.L.).

\appendix

\section{Appendix: Equivalence with non-vanishing $\e_3$ and $\e$} \label{appendix}

We extend the equivalence demonstration to the case with nonvanishing $\e_3$ and $\e$. As we remarked in the main text, these parameters capture nonlinear effect in $\D_r$. We will work perturbatively to $O(\e_3\e^0)$ and $O(\e_3^0\e)$ respectively. We begin with the FP equation. Following the same method as described for the case with $\e_3=\e=0$, we can derive the following moment equations up to $\<\D^6\>$:
\begin{align}\label{moments}
&\pd_t\<\D^2\>=2T-2m\<\D^2\>+2\e_3\<\D^2\>-2\e\<\D^4\>,\nonumber\\
&\pd_t\<\D^4\>=12T\<\D^2\>-4m\<\D^4\>-24\e_1+24\e_2\<\D^2\>\nonumber\\
&
+12\e_3\<\D^4\>-4\e\<\D^6\>,\nonumber\\
&\pd_t\<\D^6\>=30T\<\D^4\>-6m\<\D^6\>-360\e_1\<\D^2\>+120\e_2\<\D^4\>\nonumber\\
&
+30\e_3\<\D^6\>-6\e\<\D^8\>.
\end{align}
For the steady state solution approached at $t\to\infty$, we simply set the left hand side of \eqref{moments} to zero. We will solve the moments equation \eqref{moments} perturbatively by the expansion
\begin{align}
\<\D^{2n}\>=\<\D^{2n}\>_\lo+\<\D^{2n}\>_\nlo+\cdots,
\end{align}
with the leading order (LO) solution $\<\D^{2n}\>_\lo\sim O(\e_3^0\e^0)$ and the next-to-leading order (NLO) solution $\<\D^{2n}\>_\nlo$ include both $O(\e_3\e^0)$ and $O(\e_3^0\e)$. By setting $\e_3=\e=0$ in \eqref{moments}, we easily obtain the LO solution:
\begin{align}\label{lo}
&\<\D^2\>_\lo=\frac{T}{m},\quad \<\D^4\>_\lo=\frac{3T\(T^2-2m\e_1+2T\e_2\)}{m^2},\nonumber\\
&\<\D^6\>_\lo=\frac{15\(T^3+6T^2\e_2-T(6m\e_1-8\e_2^2)\)-8m\e_1\e_2}{m^3}.
\end{align}
Now we proceed to the NLO solution, which satisfies
\begin{align}
&0=-2m\<\D^2\>_\nlo+2\e_3\<\D^2\>_\lo-2\e\<\D^4\>_\lo,\nonumber\\
&0=12T\<\D^2\>_\nlo-4m\<\D^4\>_\nlo+24\e_2\<\D^2\>_\nlo\nonumber\\
&
+12\e_3\<\D^4\>_\lo-4\e\<\D^6\>_\lo.
\end{align}
Here we only keep the moments we need. The equations can be solved as
\begin{align}\label{nlo}
&\<\D^2\>_\nlo=\frac{T\e_3}{m^2}-\frac{(3T^2-6m\e_1+6T\e_2)\e}{m^3},\nonumber\\
&\<\D^4\>_\nlo=-\frac{3(-4T^2+6m\e_1-8T\e_2)\e_3}{m^3}\nonumber\\
&
-\frac{3(8T^3-36Tm\e_1+42T^2\e_2-52m\e_1\e_2+52T\e_2^2)\e}{m^4}.
\end{align}
Immediately, \eqref{lo} and \eqref{nlo} give
\begin{align}\label{nlo_c}
&\<\D^4\>^c_\lo=-\frac{6\e_1}{m}+\frac{6T\e_2}{m^2},\nonumber\\
&\<\D^4\>^c_\nlo=\frac{6(T^2-3m\e_1+4T\e_2)\e_3}{m^3}\nonumber\\
&
-\frac{6\(T^3-12Tm\e_1+15T^2\e_2-26m\e_1\e_2+26T\e_2^2\)\e}{m^4}.
\end{align}

Then we solve the Langevin equation. Similar to the simple example in the main text, we derive the following evolution equations for moments
\begin{align}\label{moments_discrete}
&\<\D_{j+1}^2\>=(1-2m\d_t)\<\D_j^2\>+(1+\e_3T^{-1}\<\D_j^2\>)\<\x_i^2\>\d_t^2\nonumber\\
&
-2\e\<\D_j^4\>\d_t,\nonumber\\
&\<\D_{j+1}^4\>=(1-4m\d_t)\<\D_j^4\>+6(\<\D_j^2\>+\e_3T^{-1}\<\D_j^4\>)\<\x_i^2\>\d_t^2\nonumber\\
&
+4\<\c_j^3\D_j\>\d_t^2+\<\h_j^4\>\d_t^4-4\e\<\D_j^6\>\d_t,\nonumber\\
&\<\D_{j+1}^6\>=(1-6m\d_t)\<\D_j^6\>+15(\<\D_j^4\>+\e_3T^{-1}\<\D_j^6\>)\<\x_j^2\>\d_t^2\nonumber\\
&
+15\<\D_j^2\>\<\h_j^4\>d_t^4+20\<\D_j^3\x_j^3\>\d_t^3-6\e\<\D_j^8\>\d_t.
\end{align}
In the above we have used the scaling properties of noises with $\d_t$ and kept only terms up to $O(\d_t)$. Treating $\<\D_j^{2n}\>$ as a continuous variables and using the expansion $\<\D_{j+1}^{2n}\>=\<\D_{j}^{2n}\>+\d_t\pd_t\<\D_{j}^{2n}\>$, we find the coefficients of $\d_t$ give nothing but the discretized version of \eqref{moments}. It follows that \eqref{moments_discrete} give the same steady state solution as \eqref{lo} and \eqref{nlo}.

\begin{figure}[htbp]
	\begin{center}
		\includegraphics[width=0.49\textwidth]{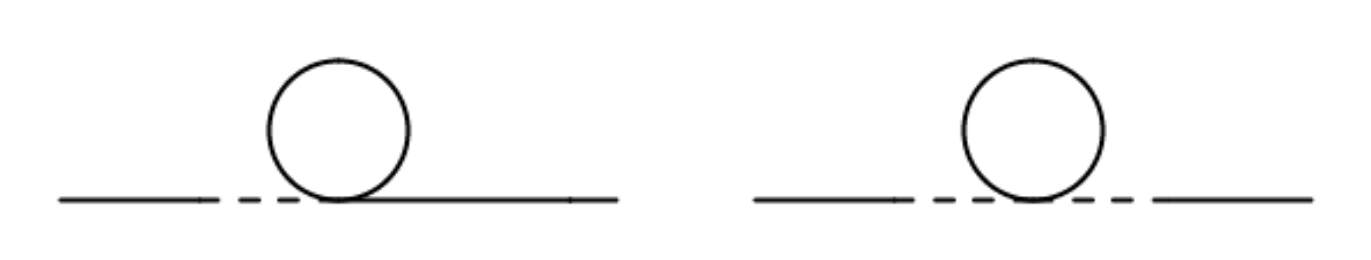}
		\caption{One-loop Feynman diagrams for $\langle \D_r(t)^2\rangle_c$.}
		\label{2pt_one_loop}
	\end{center}
\end{figure}

\begin{figure}[htbp]
	\begin{center}
		\includegraphics[width=0.49\textwidth]{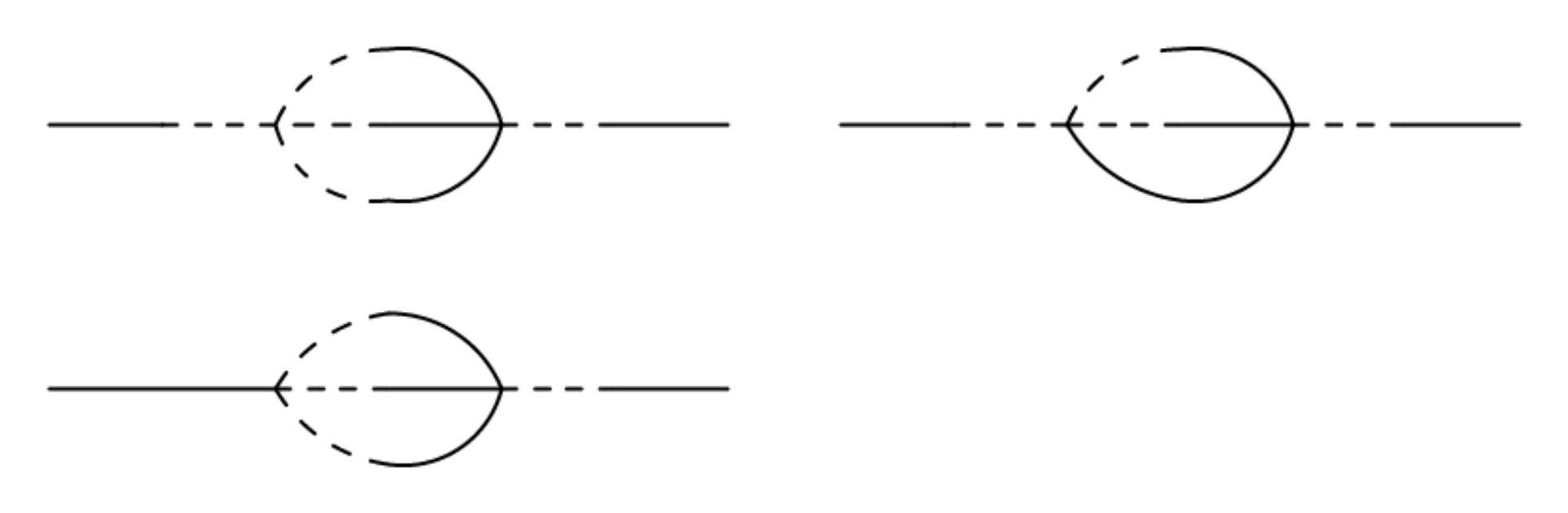}
		\caption{Two-loop Feynman diagrams for $\langle \D_r(t)^2\rangle_c$.}
		\label{2pt_two_loop}
	\end{center}
\end{figure}

\begin{figure}[htbp]
	\begin{center}
		\includegraphics[width=0.49\textwidth]{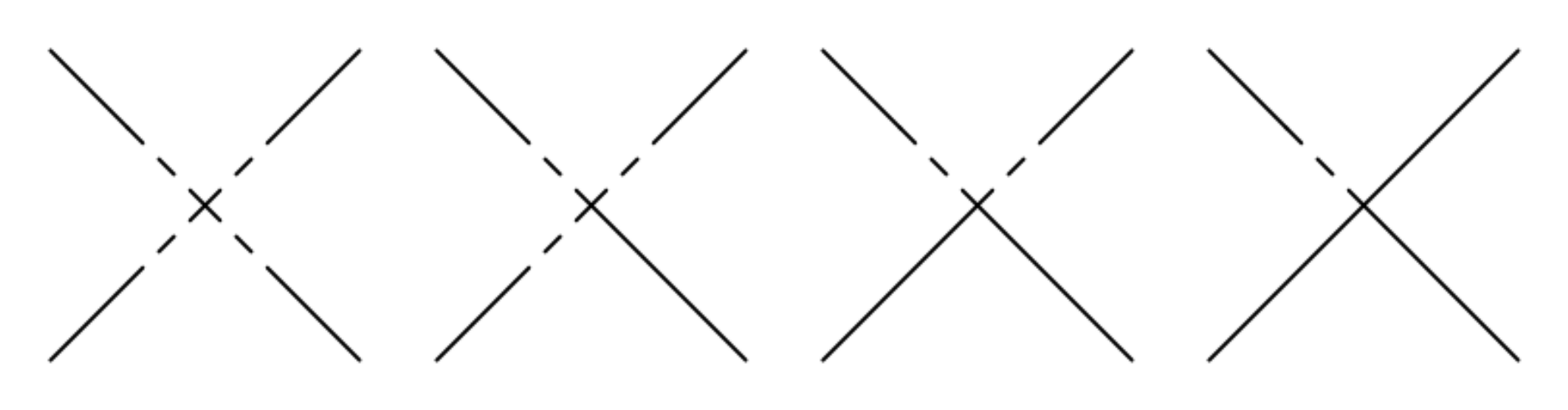}
		\caption{Tree-level Feynman diagrams for $\langle \D_r(t)^4\rangle_c$.}
		\label{4pt_tree_level}
	\end{center}
\end{figure}

\begin{figure}[htbp]
	\begin{center}
		\includegraphics[width=0.49\textwidth]{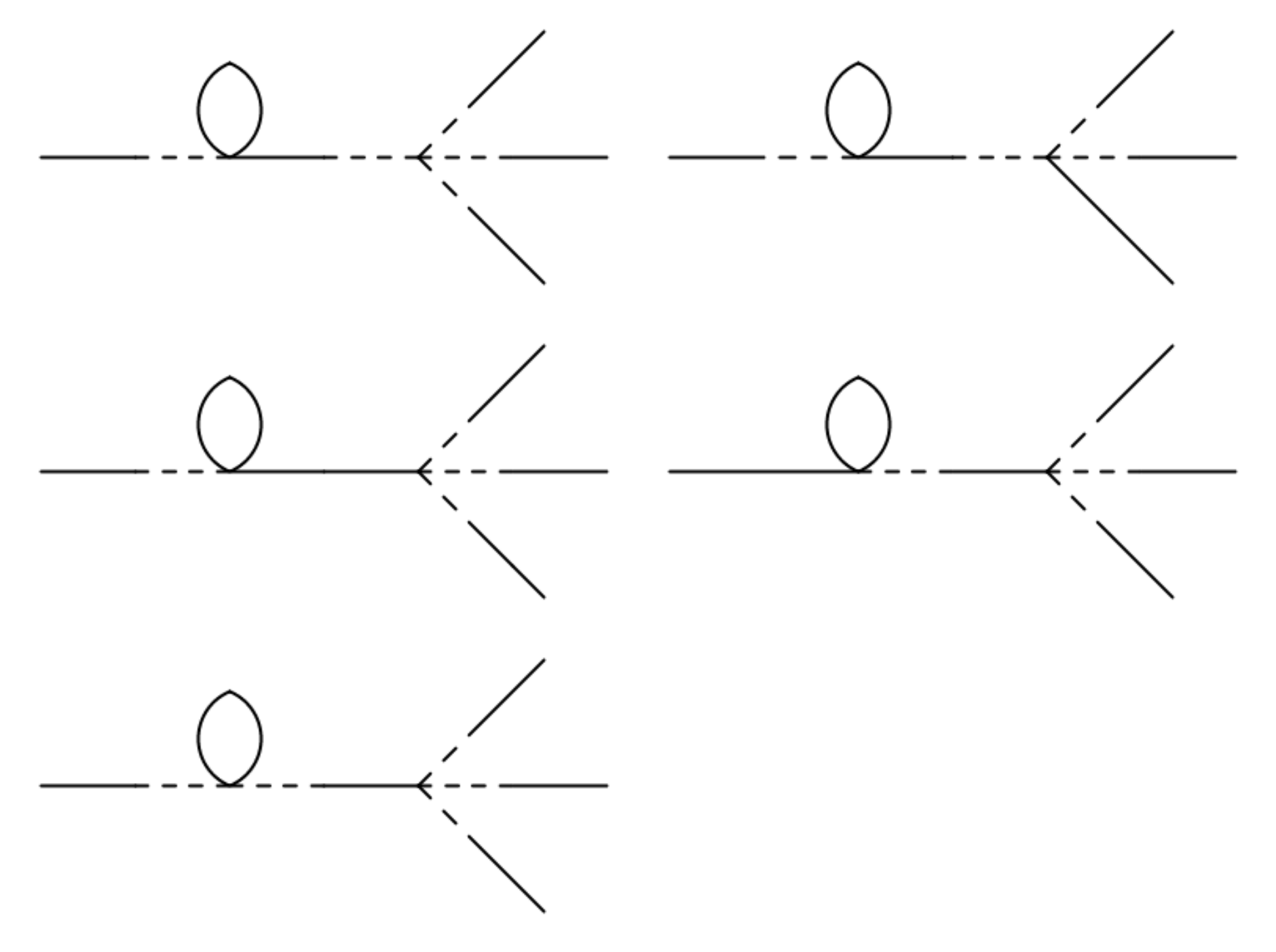}
		\caption{Reducible one-loop Feynman diagrams for $\langle \D_r(t)^4\rangle_c$.}
		\label{4pt_one_loop_red}
	\end{center}
\end{figure}

\begin{figure}[htbp]
	\begin{center}
		\includegraphics[width=0.49\textwidth]{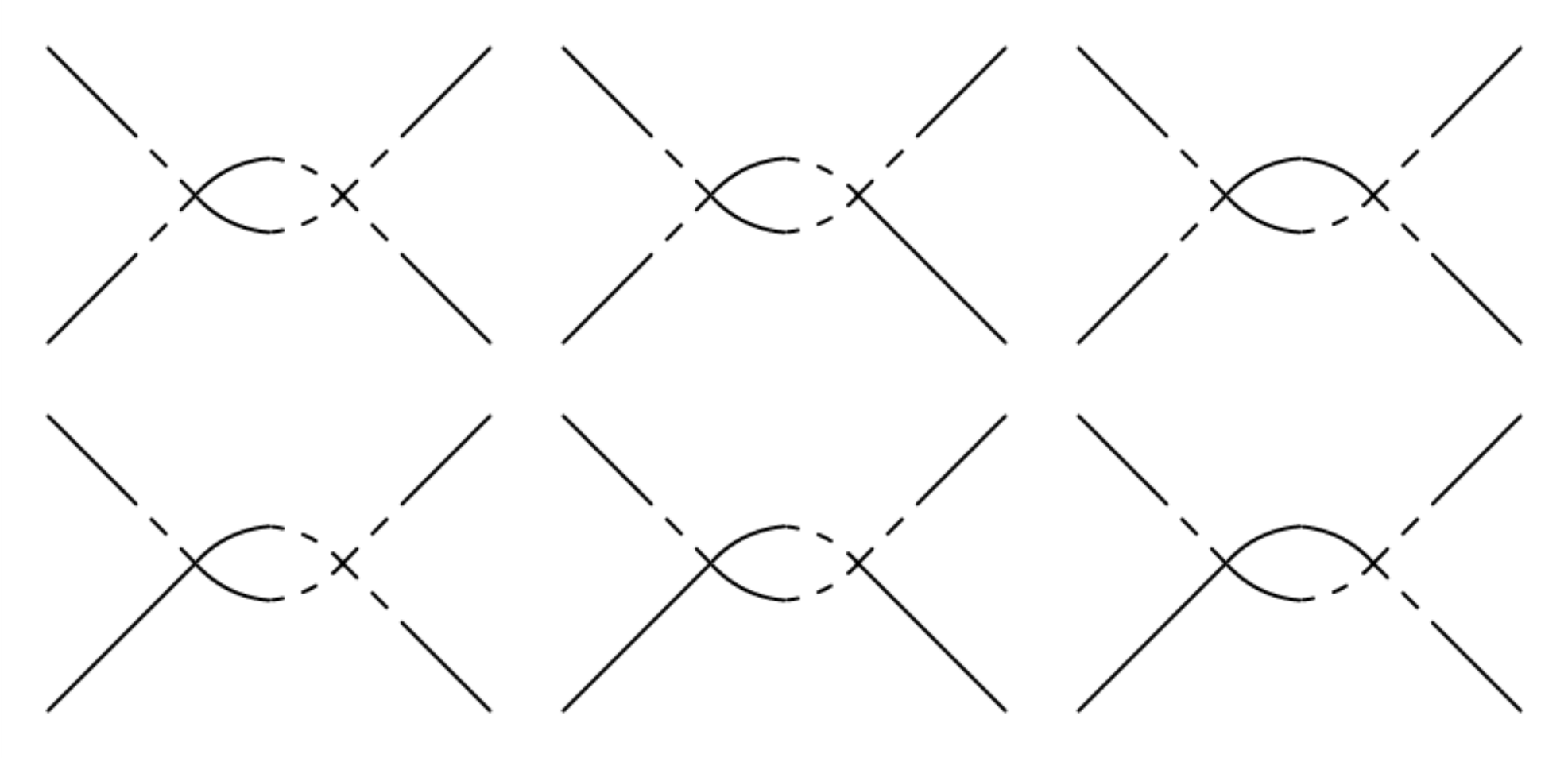}
		\caption{Irreducible one-loop Feynman diagrams for $\langle \D_r(t)^4\rangle_c$.}
		\label{4pt_one_loop_irred}
	\end{center}
\end{figure}

\begin{figure}[htbp]
	\begin{center}
		\includegraphics[width=0.49\textwidth]{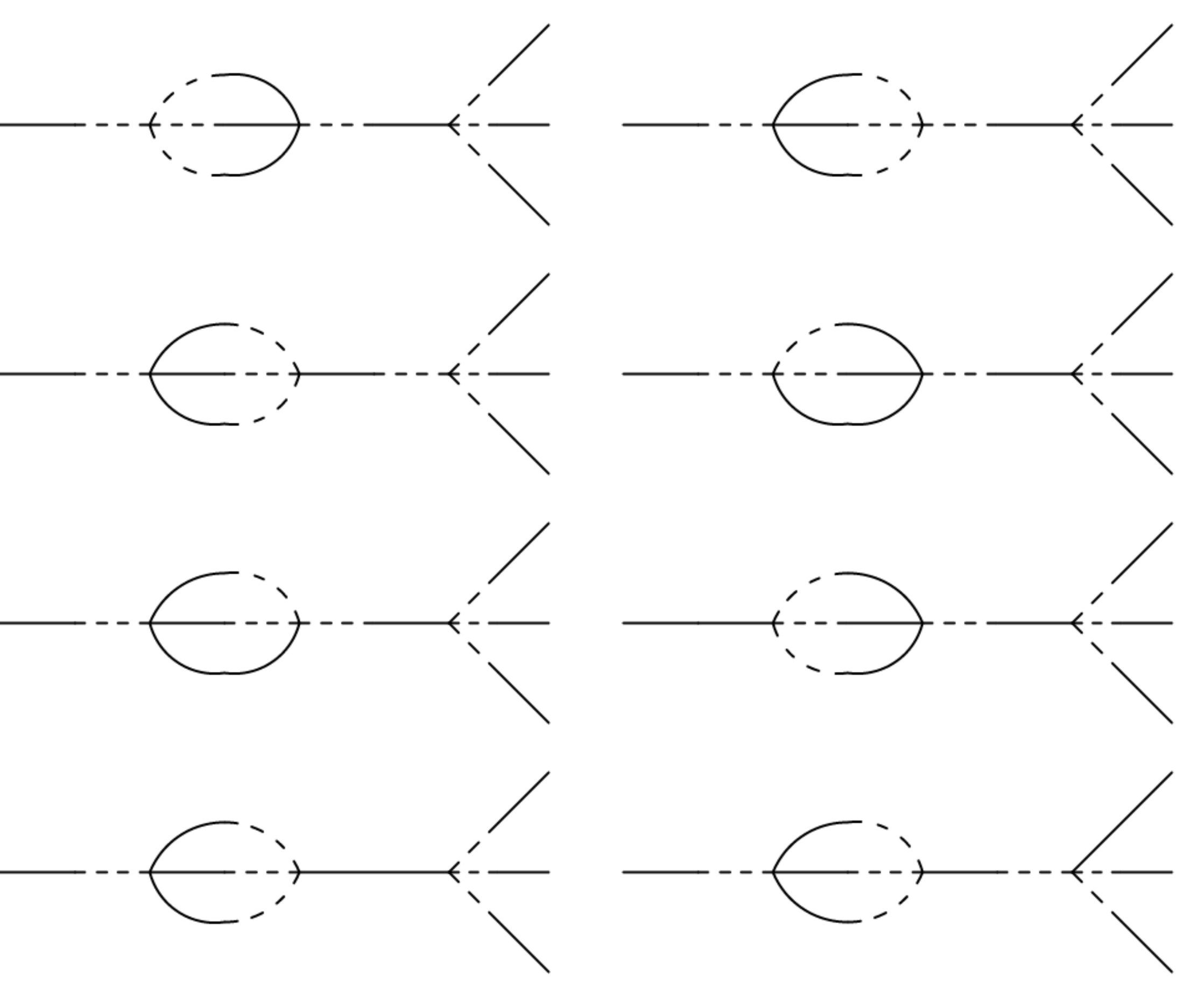}
		\caption{Reducible two-loop Feynman diagrams for $\langle \D_r(t)^4\rangle_c$.}
		\label{4pt_two_loop_red}
	\end{center}
\end{figure}

\begin{figure}[htbp]
	\begin{center}
		\includegraphics[width=0.49\textwidth]{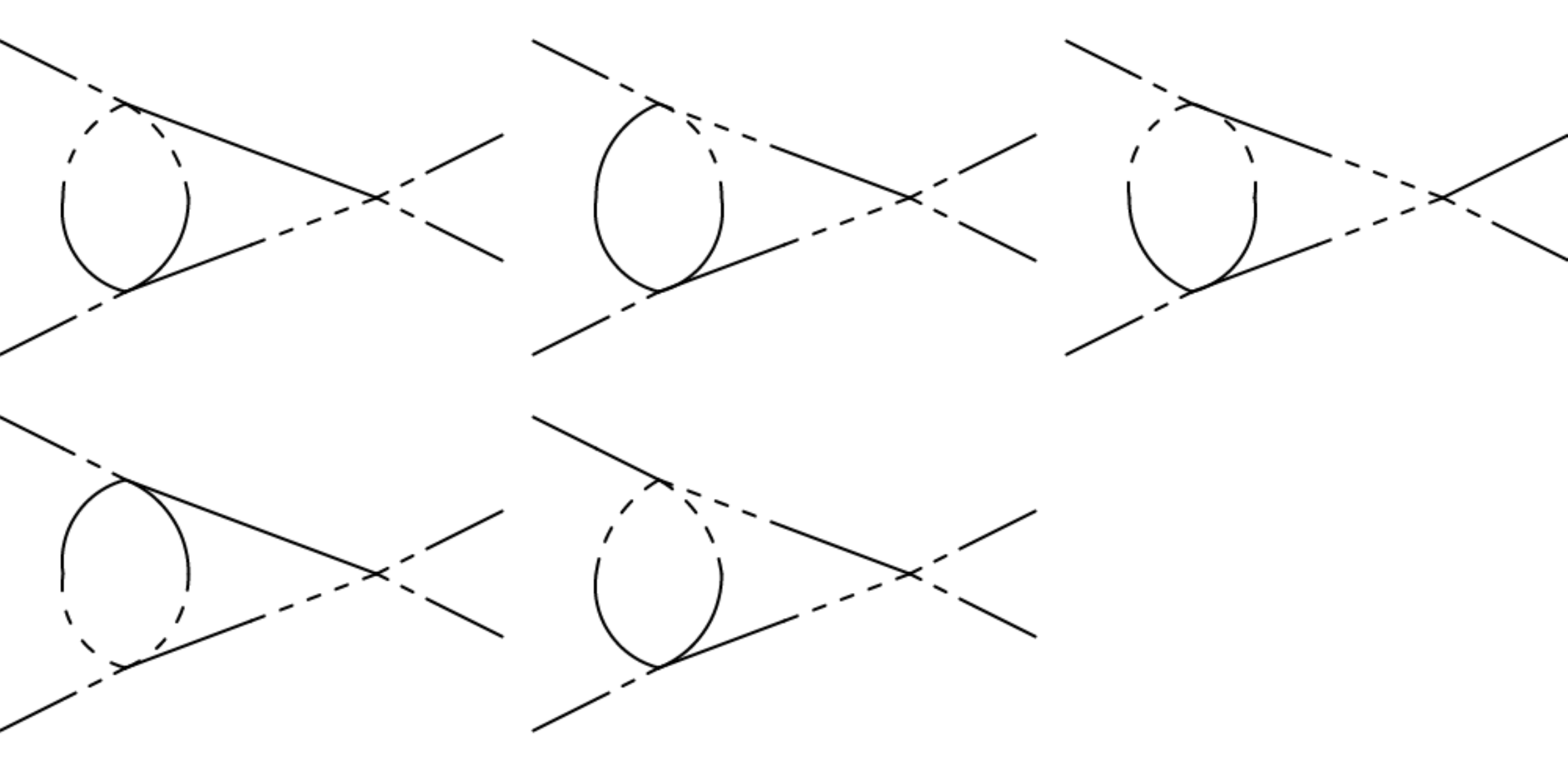}
		\caption{Irreducible two-loop Feynman diagrams for $\langle \D_r(t)^4\rangle_c$.}
		\label{4pt_two_loop_irred}
	\end{center}
\end{figure}

Finally, we turn to diagrammatic computations of $\<\D(t)^2\>$ and $\<\D(t)^4\>_c$. The LO results have been obtained in the main text. For the NLO results, we need diagrams with one vertex of either $aarr$ type or $aaar$ type and arbitrary number of other vertices. We first look at $\<\D(t)^2\>_\nlo$, which receives contributions from one-loop and two-loop diagrams shown respectively in Fig.~\ref{2pt_one_loop} and Fig.~\ref{2pt_two_loop}.
The evaluations of them are straightforward. We obtain in the end
\begin{align}\label{2pt}
\<\D(t)^2\>_\nlo=-\frac{3T^2\e}{m^3}+\frac{T\e_3}{m^2}+\frac{6\e_1\e}{m^2}-\frac{6T\e_2\e}{m^3},
\end{align}
which is in perfect agreement with \eqref{nlo}. The situation of $\<\D(t)^4\>_\nlo^c$ is more complicated: it receive contributions from both tree-level diagrams and loop diagrams; moreover, the latter contain both reducible and irreducible ones. The full tree-level diagrams are shown in Fig.~\ref{4pt_tree_level}.
The first two are shown already in Fig.~$3$ of the main text and give the LO result ($27$) of the main text. The last two are easily evaluated to give the NLO result
\begin{align}\label{4pt_tree}
\<\D(t)^4\>^{c, \rm tree}_\nlo=\frac{6T^2\e_3}{m^3}-\frac{6T^3\e}{m^4}.
\end{align}
The reducible and irreducible one-loop diagrams are shown respectively in Fig.~\ref{4pt_one_loop_red} and \ref{4pt_one_loop_irred}.
The reducible diagrams (Fig~\ref{4pt_one_loop_red}) are evaluated to give
\begin{align}\label{4pt_1loop_red}
\<\D(t)^4\>^{c,\rm 1loop-red}_\nlo = \frac{18T\e_1\e}{m^3}-\frac{36T^2\e_2\e}{m^4}+\frac{6T\e_2\e_3}{m^3}.
\end{align}
The irreducible diagrams (Fig.~\ref{4pt_one_loop_irred}) give rise to the following results
\begin{align}\label{4pt_1loop_irred}
&\<\D(t)^4\>^{c,\rm 1loop-irred}_\nlo=-\frac{18\e_1\e_3}{m^2}+\frac{18T\e_2\e_3}{m^3}+\frac{54T\e_1\e}{m^3}\nonumber\\
&-\frac{54T^2\e_2\e}{m^4}.
\end{align}
At two-loop level, the reducible and irreducible diagrams are shown respectively in Fig.~\ref{4pt_two_loop_red} and \ref{4pt_two_loop_irred}.
Two-loop irreducible diagrams of a different topology are excluded by the presence of loop containing only $ra$ type propagators. Being careful with the combinatoric factors, we arrive at
\begin{align}\label{4pt_2loop_red}
\<\D(t)^4\>^{c,\rm 2loop-red}_\nlo=\frac{48\e_1\e_2\e}{m^3}-\frac{48T\e_2^2\e}{m^4}.
\end{align}
for the reducible diagrams (Fig.~\ref{4pt_two_loop_red}) and
\begin{align}\label{4pt_2loop_irred}
\<\D(t)^4\>^{c,\rm 2loop-irred}_\nlo=\frac{108\e_1\e_2\e}{m^3}-\frac{108T\e_2^2\e}{m^4},
\end{align}
for the irreducible diagrams (Fig.~\ref{4pt_two_loop_irred}). The final result for $\<\D(t)^4\>^c_\nlo$ is the sum of \eqref{4pt_tree}, \eqref{4pt_1loop_red}, \eqref{4pt_1loop_irred}, \eqref{4pt_2loop_red} and \eqref{4pt_2loop_irred}, which reads
\begin{align}\label{4pt}
&\<\D(t)^4\>^c_\nlo=\frac{6T^2\e_3}{m^3}-\frac{6T^3\e}{m^4}+\frac{72T\e_1\e}{m^3}-\frac{18\e_1\e_3}{m^2}\nonumber\\
&+\frac{24T\e_2\e_3}{m^3}
-\frac{90T^2\e_2\e}{m^4}+\frac{156\e_1\e_2\e}{m^3}-\frac{156T\e_2^2\e}{m^4}.
\end{align}
This is also in full agreement with \eqref{nlo_c}. The agreement serves as a nontrivial demonstration of the equivalence among the three formulations.

\bibliography{EFT}



\end{document}